\documentclass[11pt]{article}

\usepackage{color}

\usepackage{greekbf}

\usepackage{amsfonts,amssymb,amsthm}
\usepackage{amsmath,amssymb}
\usepackage{mathrsfs}
\usepackage{enumerate}
\usepackage{srcltx}
\usepackage[ansinew]{inputenc}
\usepackage{mathtools}
    \mathtoolsset{showonlyrefs}
    \mathtoolsset{centercolon}
\usepackage{graphicx}
 \usepackage{hyperref}

\DeclareMathAlphabet{\mathbf}{OT1}{cmr}{bx}{it}
\DeclareMathAlphabet{\mathssb}{OT1}{cmss}{bx}{n}
\DeclareMathAlphabet{\mathssn}{OT1}{cmss}{m}{n}
\DeclareMathAlphabet{\mathub}{OT1}{cmr}{b}{n}

\DeclareMathOperator{\dr}{d\!}

\DeclareMathOperator{\A}{\mathcal{A}}

\begin{document}

\begin{center}

\end{center}
%
% Macro per LaTeX
%
%    aggiornati al 20-set-99
%
%====  CARATTERI SPECIALI ==========================
%
\newcommand{\bydef}{\,\raise.050ex\hbox{\rm:}\kern-.025em\hbox{\rm=}\,}
\newcommand{\defby}{=\raise.075ex\hbox{\kern-.325em\hbox{\rm:}}\,}
\newcommand{\mtrp}  {{-\!\top}} % -trasposto alla "adc"
\newcommand{\bdot}  {{\scriptscriptstyle\bullet}}
\def\qed{\relax\ifmmode\hskip2em \Box\else\unskip\nobreak\hskip1em $\Box$\fi}
%
%==== VARIE =====================================
%
%\newcommand {\dim}[1] {\mathtt{dim}(#1)}
\newcommand {\eps} {\varepsilon} % espilon
\newcommand {\vp} {\varphi}      % phi
\newcommand {\0} {\textbf{0}}    % 0 bold
\newcommand {\1} {\textbf{1}}    % 0 bold
%
%==== CALLIGRAFICI ==================
\newcommand {\Ac}  {\mathcal{A}}
\newcommand {\Bc}  {\mathcal{B}}
\newcommand {\Cc}  {\mathcal{C}}
\newcommand {\Dc}  {\mathcal{D}}
\newcommand {\Ec}  {\mathcal{E}}
\newcommand {\Fc}  {\mathcal{F}}
\newcommand {\Gc}  {\mathcal{G}}
\newcommand {\Hc}  {\mathcal{H}}
\newcommand {\Kc}  {\mathcal{K}}
\newcommand {\Ic}  {\mathcal{I}}
\newcommand {\Jc}  {\mathcal{J}}
\newcommand {\Lc}  {\mathcal{L}}
\newcommand {\Mc}  {\mathcal{M}}
\newcommand {\Nc}  {\mathcal{N}}
\newcommand {\Oc}  {\mathcal{O}}
\newcommand {\Pc}  {\mathcal{P}}
\newcommand {\Rc}  {\mathcal{R}}
\newcommand {\Sc}  {\mathcal{S}}
\newcommand {\Tc}  {\mathcal{T}}
\newcommand {\Uc}  {\mathcal{U}}
\newcommand {\Vc}  {\mathcal{V}}
\newcommand {\Wc}  {\mathcal{W}}
\newcommand {\Zc}  {\mathcal{Z}}
%
%==== BOLD ====
%
\newcommand {\ab} {\mathbf{a}}
\newcommand {\bb} {\mathbf{b}}
\newcommand {\cb} {\mathbf{c}}
\newcommand {\db} {\mathbf{d}}
\newcommand {\eb} {\mathbf{e}}
\newcommand {\fb} {\mathbf{f}}
\newcommand {\gb} {\mathbf{g}}
\newcommand {\hb} {\mathbf{h}}
\newcommand {\ib} {\mathbf{i}}
\newcommand {\kb} {\mathbf{k}}
\newcommand {\lb} {\mathbf{l}}
\newcommand {\mb} {\mathbf{m}}
\newcommand {\nb} {\mathbf{n}}
\newcommand {\pb} {\mathbf{p}}
\newcommand {\qb} {\mathbf{q}}
\newcommand {\rb} {\mathbf{r}}
\renewcommand {\sb} {\mathbf{s}}
\newcommand {\tb} {\mathbf{t}}
\newcommand {\xb} {\mathbf{x}}
\newcommand {\yb} {\mathbf{y}}
\newcommand {\ub} {\mathbf{u}}
\newcommand {\vb} {\mathbf{v}}
\newcommand {\wb} {\mathbf{w}}
\newcommand {\zb} {\mathbf{z}}
\newcommand {\Ab} {\mathbf{A}}
\newcommand {\Bb} {\mathbf{B}}
\newcommand {\Cb} {\mathbf{C}}
\newcommand {\Db} {\mathbf{D}}
\newcommand {\Eb} {\mathbf{E}}
\newcommand {\Fb} {\mathbf{F}}
\newcommand {\Gb} {\mathbf{G}}
\newcommand {\Hb} {\mathbf{H}}
\newcommand {\Kb} {\mathbf{K}}
\newcommand {\Jb} {\mathbf{J}}
\newcommand {\Ib} {\mathbf{I}}
\newcommand {\Lb} {\mathbf{L}}
\newcommand {\Mb} {\mathbf{M}}
\newcommand {\Nb} {\mathbf{N}}
\newcommand {\Ob} {\mathbf{O}}
\newcommand {\Pb} {\mathbf{P}}
\newcommand {\Qb} {\mathbf{Q}}
\newcommand {\Rb} {\mathbf{R}}
\newcommand {\Sb} {\mathbf{S}}
\newcommand {\Tb} {\mathbf{T}}
\newcommand {\Vb} {\mathbf{V}}
\newcommand {\Wb} {\mathbf{W}}
\newcommand {\Xb} {\mathbf{X}}
\newcommand {\Zb} {\mathbf{Z}}
%

%
%==== ASSI (minuscole) e PUNTI (maiuscole) dello spazio euclideo ====
%
\newcommand {\ax} {\mathrm{a}}
\newcommand {\bx} {\mathrm{b}}
\newcommand {\cx} {\mathrm{c}}
\newcommand {\dx} {\mathrm{d}}
\newcommand {\ex} {\mathrm{e}}
\newcommand {\fx} {\mathrm{f}}
\newcommand {\gx} {\mathrm{g}}
\newcommand {\hx} {\mathrm{h}}
\newcommand {\kx} {\mathrm{k}}
\newcommand {\lx} {\mathrm{l}}
\newcommand {\mx} {\mathrm{m}}
\newcommand {\nx} {\mathrm{n}}
\newcommand {\ox} {\mathrm{o}}
\newcommand {\px} {\mathrm{p}}
\newcommand {\qx} {\mathrm{q}}
\newcommand {\rx} {\mathrm{r}}
\newcommand {\sx} {\mathrm{s}}
\newcommand {\tx} {\mathrm{t}}
\newcommand {\xx} {\mathrm{x}}
\newcommand {\yx} {\mathrm{y}}
\newcommand {\wx} {\mathrm{w}}
\newcommand {\ux} {\mathrm{u}}
\newcommand {\vx} {\mathrm{v}}
\newcommand {\zx} {\mathrm{z}}
\newcommand {\Ax} {\mathrm{A}}
\newcommand {\Bx} {\mathrm{B}}
\newcommand {\Cx} {\mathrm{C}}
\newcommand {\Dx} {\mathrm{D}}
\newcommand {\Ex} {\mathrm{E}}
\newcommand {\Fx} {\mathrm{F}}
\newcommand {\Gx} {\mathrm{G}}
\newcommand {\Hx} {\mathrm{H}}
\newcommand {\Kx} {\mathrm{K}}
\newcommand {\Jx} {\mathrm{J}}
\newcommand {\Ix} {\mathrm{I}}
\newcommand {\Lx} {\mathrm{L}}
\newcommand {\Mx} {\mathrm{M}}
\newcommand {\Nx} {\mathrm{N}}
\newcommand {\Ox} {\mathrm{O}}
\newcommand {\Px} {\mathrm{P}}
\newcommand {\Qx} {\mathrm{Q}}
\newcommand {\Rx} {\mathrm{R}}
\newcommand {\Sx} {\mathrm{S}}
\newcommand {\Tx} {\mathrm{T}}
\newcommand {\Vx} {\mathrm{V}}
\newcommand {\Wx} {\mathrm{W}}
\newcommand {\Xx} {\mathrm{X}}
\newcommand {\Yx} {\mathrm{Y}}
\newcommand {\Zx} {\mathrm{Z}}
%

%
%==== ORLATI =============
%
\newcommand {\Real} {\mathbb{R}}
\newcommand {\Aos} {\mbox{$\scriptstyle\mathbb{A}$}}
\newcommand {\Ao} {\mathbb{A}}
\newcommand {\Bo} {\mathbb{B}}
\newcommand {\Co} {\mathbb{C}}
\newcommand {\Cop} {\mbox{$\scriptstyle\mathbb{C}$}}
\newcommand {\Do} {\mathbb{D}}
\newcommand {\Fo} {\mathbb{F}}
\newcommand {\Go} {\mathbb{G}}
\newcommand {\Io} {\mathbb{I}}
\newcommand {\Mo} {\mathbb{M}}
\newcommand {\Ko} {\mathbb{K}}
\newcommand {\No} {\mathbb{N}}
\newcommand {\Ro} {\mathbb{R}}
\newcommand {\So} {\mathbb{S}}
\newcommand {\To} {\mathbb{T}}
\newcommand {\Vo} {\mathbb{V}}
\newcommand {\Zo} {\mathbb{Z}}
%
%==== EULER SCRIPT =============
% non funzionano! perche?
%\usepackage{euler}
\newcommand {\Xes} {\mathscr{X}}
\newcommand {\ges} {\mathscr{g}}
\newcommand {\wes} {\mathscr{W}}
%
%==== EULER FRACTUR =============
%
\newcommand {\aef} {\mathfrak{a}}
\newcommand {\fef} {\mathfrak{f}}
\newcommand {\gef} {\mathfrak{g}}
\newcommand {\hef} {\mathfrak{h}}
\newcommand {\mef} {\mathfrak{m}}
\newcommand {\nef} {\mathfrak{n}}
\newcommand {\kef} {\mathfrak{k}}
\newcommand {\wef} {\mathfrak{w}}
\newcommand {\sef} {\mathfrak{s}}
\newcommand {\zef} {\mathfrak{z}}
\newcommand {\Def} {\mathfrak{D}}
\newcommand {\Fef} {\mathfrak{F}}
\newcommand {\Mef} {\mathfrak{M}}
\newcommand {\Nef} {\mathfrak{N}}
\newcommand {\Ref} {\mathfrak{R}}
\newcommand {\Sef} {\mathfrak{S}}
\newcommand {\Xef} {\mathfrak{X}}
%
%==== TIPOGRAFICI =============
%
\newcommand {\att} {\mathtt{a}}
\newcommand {\btt} {\mathtt{b}}
\newcommand {\ctt} {\mathtt{c}}
\newcommand {\dtt} {\mathtt{d}}
\newcommand {\ftt} {\mathtt{f}}
\newcommand {\gtt} {\mathtt{g}}
\newcommand {\mtt} {\mathtt{m}}
\newcommand {\ntt} {\mathtt{n}}
\newcommand {\htt} {\mathtt{h}}
\newcommand {\ptt} {\mathtt{p}}
\newcommand {\qtt} {\mathtt{q}}
\newcommand {\rtt} {\mathtt{r}}
\newcommand {\stt} {\mathtt{s}}
\newcommand {\ttt} {\mathtt{t}}
\newcommand {\vtt} {\mathtt{v}}
\newcommand {\wtt} {\mathtt{w}}
\newcommand {\ztt} {\mathtt{z}}
\newcommand {\Ftt} {\mathtt{F}}
\newcommand {\Stt} {\mathtt{P}}
\newcommand {\Wtt} {\mathtt{W}}
%
%
%======Lowercase greek letters BOLD============
%
%
%\def\nub{\hbox{\mbo {\char 23}}}%="0?17
%\def\xib{\hbox{\mbo {\char 24}}}%="0?18
%\def\pib{\hbox{\mbo {\char 25}}}%="0?19
%\def\rhob{\hbox{\mbo {\char 26}}}%="0?1A
%\def\sigmab{\hbox{\mbo {\char 27}}}%="0?1B
\newcommand {\alfab}     {\mathbf{\alpha}}
\newcommand {\betab}     {\mathbf{\beta}}
\newcommand {\gammab}    {\mathbf{\gamma}}
\newcommand {\deltab}    {\mathbf{\delta}}
\newcommand {\epsilonb}  {\mathbf{\epsilon}}
\newcommand {\epsb}      {\mathbf{\varepsilon}}
\newcommand {\zetab}     {\mathbf{\zeta}}
\newcommand {\etab}      {\mathbf{\eta}}
\newcommand {\tetab}     {\mathbf{\teta}}
\newcommand {\vtetab}    {\mathbf{\vartheta}}
\newcommand {\iotab}     {\mathbf{\iota}}
\newcommand {\kappab}    {\mathbf{\kappa}}
\newcommand {\lambdab}   {\mathbf{\lambda}}
\newcommand {\mub}       {\mathbf{\mu}}
\font\mbo=cmmib10 scaled \magstephalf
\newcommand{\nub}   {\hbox{\mbo {\char 23}}}%="0?17
\newcommand {\csib}      {\mathbf{\xi}}
\newcommand {\xib}      {\mathbf{\xi}}
\newcommand {\pib}       {\mathbf{\pi}}
\newcommand {\varrhob}   {\mathbf{\varrho}}
\newcommand {\sigmab}    {\hbox{\mbo {\char 27}}}
\newcommand{\taub}   {\hbox{\mbo {\char 28}}}%="0?17
\newcommand {\upsilonb}  {\mathbf{\upsilon}}
\newcommand {\phib}      {\mathbf{\phi}}
\newcommand {\varphib}   {\mathbf{\varphi}}
\newcommand {\chib}      {\mathbf{\chi}}
\newcommand {\psib}       {\mathbf{\psi}}
\newcommand {\omegab}    {\mathbf{\omega}}
%
%
%======Uppercase greek letters BOLD============
%
%
\newcommand {\Gammab}    {\mathbf{\Gamma}}
\newcommand {\Deltab}    {\mathbf{\Delta}}
\newcommand {\Tetab}     {\mathbf{\Theta}}
\newcommand {\Lambdab}   {\mathbf{\Lambda}}
\newcommand {\Csib}      {\mathbf{\Xi}}
\newcommand {\Pib}       {\mathbf{\Pi}}
\newcommand {\Sigmab}    {\mathbf{\Sigma}}
\newcommand {\Phib}      {\mathbf{\Phi}}
\newcommand {\Psib}      {\mathbf{\Psi}}
\newcommand {\Omegab}    {\mathbf{\Omega}}
%
%==== LIN & COMPANY ================
%
\newcommand {\Lin} {\mathbb{L}\mathtt{in}}
\newcommand {\Sym} {\mathbb{S}\mathtt{ym}}
\newcommand {\Psym} {\mathbb{PS}\mathtt{ym}}
\newcommand {\Skw} {\mathbb{S}\mathtt{kw}}
\newcommand {\SO} {\mathcal{SO}}
\newcommand {\GL} {\mathcal{G}l}
\newcommand {\Rot} {\mathbb{R}\mathtt{ot}}
%
%==== OPERATORI ================
%
\newcommand {\tr}[1]{\mbox{tr}\, #1}
\newcommand {\psym} {\mbox{sym}}
\newcommand {\pskw} {\mbox{skw}}
\newcommand {\win}[2] {( #1 \cdot #2 )_\wedge }
\newcommand {\modulo}[1] {\left|#1\right|}
\newcommand {\sph} {\mbox{sph}}
\newcommand {\dev} {\mbox{dev}}
\newcommand {\sgn} {\mbox{sgn}}
\newcommand {\lin} {\mbox{Lin}}
%
%==== OPERATORI DIFFERENZIALI ================
%
\newcommand{\dvg} {\mathrm{div}\,}
\newcommand{\dvgt} {\mathrm{d{\widetilde i}v}\,}    % divergenza
\newcommand{\grd} {\mathrm{grad}\,}    % gradiente
\newcommand{\Grd} {\mathrm{Grad}\,}    % Gradiente
\newcommand{\dl}  {\delta}             % delta
\def\gradtwo{\mathord{\nabla^{\scriptscriptstyle(2)}}}
\def\Div{\mathop{\hbox{Div}}}
\def\div{\mathop{\hbox{div}}}
\def\mis{\mathop{\hbox{mis}}}
\def\eps{\varepsilon}
%
%==== MISCELLANEA ================
%

\newcommand\ph{\varphi}
\newcommand{\adj} {\att\dtt}     % aggiunto piccolo

\newcommand{\va}{\mathbf{a}}
\newcommand{\vn}{\mathbf{\nu}}
\newcommand{\vt}{\mathbf{\tau}}
\newcommand{\dn}{\partial_{\mathbf{\nu}}}
\newcommand{\dt}{\partial_{\mathbf{\tau}}}
\newcommand{\ord}{\scriptscriptstyle}
\newcommand{\tD}{\mathbf{E}}  %  tensore della deformazione
\newcommand{\tS}{\mathbf{S}}  %  tensore di sforzo
\newcommand{\tE}{\mathbb{C}}
\newcommand{\tPf}{\mathbb{P}}
\newcommand{\tPr}{\mathbf{P}}
\newcommand{\tC}{\mathbf{C}}
\newcommand{\tP}{{\scriptstyle\mathbb{C}}}   %  tensore piezoelettrico
\newcommand{\tDl}{\mathbf{C}}      %  tensore dielettrico

\newcommand{\Cuno}{\mathbf{c}_1}
\newcommand{\Cdue}{\mathbf{c}_2}
\newcommand{\veralf}{\mathbf{c}_\alpha} %  C1 e C2 in alfa
\newcommand{\verbet}{\mathbf{c}_\beta} %  C1 e C2 in alfa
\newcommand{\vz}{\mathbf{z}}
\newcommand{\sym}{\mathop{\mathrm{sym}}}

\newcommand{\f}{f}
\newcommand{\g}{g}
\newcommand{\h}{h}
\newcommand{\w}{w}
\renewcommand{\l}{l}

\renewcommand{\r}{r}
\newcommand{\s}{s}

\newcommand{\autof}{\mathrm{\skew 0\overline{w}}}
\newcommand{\W}{W}

\newcommand{\df}{f'}
\newcommand{\dg}{g'}
\renewcommand{\dh}{h'}
\newcommand{\ddf}{f''}
\newcommand{\ddg}{g''}
\newcommand{\ddh}{h''}

\newcommand\modv[1]{|{#1}|}

\newcommand\arr[1]{\overrightarrow{#1}}
\newcommand{\cartref}{\{O;x_1,x_2,x_3\}}
\newcommand{\orthframe}{({\bf e}_1, {\bf e}_2, {\bf e}_3)}


\newcommand\email[1]{\texttt{#1}}
\newcommand\at{:}

\begin{center}
 {\bf \Large
On shear and torsion factors in the theory of linearly elastic
rods}
\end{center}
\medskip

\begin{center}
{\large Antonino Favata \quad
        Andrea Micheletti \quad
        Paolo Podio-Guidugli
}\end{center}

\begin{center}
 \noindent Dipartimento di Ingegneria Civile, Universit\`a di Roma Tor Vergata\footnote{Via Politecnico 1, 00133 Rome, Italy. \\
{\null} \quad \ Email:
\begin{minipage}[t]{30em}
\email{favata@ing.uniroma2.it} (A. Favata)\\
\email{micheletti@ing.uniroma2.it} (A. Micheletti)\\
\email{ppg@uniroma2.it} (P. Podio-Guidugli)
\end{minipage}}
\small
\end{center}
\medskip

\begin{abstract}
\noindent {\footnotesize Lower bounds for the factors entering the
standard notions of shear and torsion stiffness for a linearly
elastic rod are established in a new and simple way. The proofs
are based on the following criterion to identify the stiffness
parameters entering rod theory: the rod's stored-energy density
per unit length expressed in terms of force and moment resultants
should equal the stored-energy density per unit length expressed
in terms of stress components of a Saint--Venant cylinder subject
to either flexure or torsion, according to the case. It is shown
that the shear factor is always greater than one, whatever the
cross section, a fact that is customarily stated without proof in
textbooks of structure mechanics; and that the torsion factor is
also greater than one, except when the cross section is a circle
or a circular annulus, a fact that is usually proved making use of
Saint--Venant's solution in terms of displacement components.
\medskip

}
\end{abstract}

\section{Introduction}

\begin{figure}[h]
\centering
\includegraphics[scale=0.9]{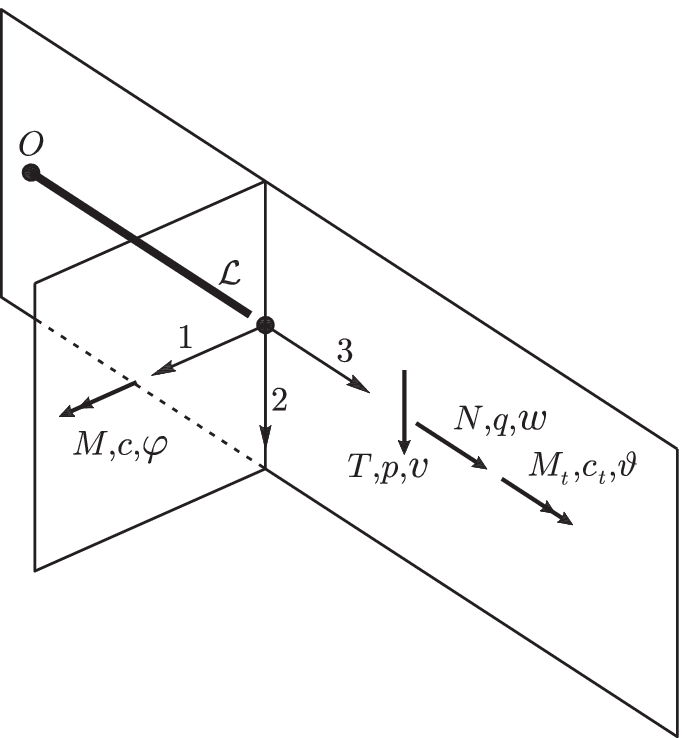}
%\picplace{5cm}{2cm} % Give the correct figure height and width in cm
%
\caption{}
\label{uno} % Give a unique label
\end{figure}

When a direct approach is chosen to expound the standard theory of
linearly elastic rods with a straight axis $\mathcal L$, the
\emph{Principle of Virtual Working} is laid down:
\begin{equation}\label{virt}
\int_{\mathcal L}(T\gamma+N\varepsilon+M\psi+M_t\psi_t)dx_3
=:{\mathcal W}^{i}={\mathcal W}^e:=\int_{\mathcal L}(p v +q w +c
\varphi + c_t \vartheta )dx_3\,.
\end{equation}
The \emph{internal working} ${\mathcal W}^{i}$ collects four pairs
$(T,\gamma)$, \dots , $(M_t,\psi_t)$ of kinematic and dynamic
fields in duality, with $T,N,M,$ and $M_t$ the \emph{shear force,
normal force, bending moment} and \emph{twisting moment} and with
$\gamma,\varepsilon,\psi$ and $\psi_t$ the \emph{shear, extension,
flexion} and \emph{torsion measures}. The geometric compatibility
conditions are:
$$\gamma = v^\prime +\varphi, \quad \varepsilon=w^\prime, \quad\psi=\varphi^\prime\;\;\textrm{and}\;\;\psi_t=\vartheta^\prime,
$$
where $v$ and $w$ are the \emph{transverse} and \emph{axial
displacements}, $\varphi$ and $\vartheta$ are the rotations about,
respectively, the $x_2$-axis and the $x_3$-axis (a superscript
prime denotes differentiation with respect to $x_3$; see Fig.
\ref{uno});
the  \emph{external working} ${\mathcal W}^{e}$ sets these
kinematic variables in duality with the applied loads per unit
length $p,q,c$ and $c_t$.\footnote{To keep things simple, in
laying down (\refeq{virt}) we have ignored  the contributions to
external and internal workings of concentrated loads, axial
pin-junctions \emph{et similia}.} The constitutive equations are:
\begin{equation}\label{roden}
T=\mathfrak{s}_s\gamma,\ldots,M_t=\mathfrak{s}_t\psi_t,
\end{equation}
all the \emph{stiffness moduli}
$\mathfrak{s}_s,\ldots,\mathfrak{s}_t$ being assumed positive; the
\emph{stored-energy density} per unit length is:
\begin{equation}\label{stenrod}
\frac{1}{2}\,\frac{T^2}{\mathfrak{s}_s}+\ldots+\frac{1}{2}\,\frac{M_t^2}{\mathfrak{s}_t}
\end{equation}
(see, e.g., \cite{GeT}).

To make use of such a theory --- in fact, to make use of any
structure theory formulated by way of a direct approach --- the
issue of \emph{parameter identification} must be dealt with. In
particular, one would like to gather from some suitable
equilibrium theory  for a rod-like three-dimensional body subject
to flexure and torsion enough information to choose the `right'
shear and torsion stiffness parameters
$\mathfrak{s}_s,\mathfrak{s}_t$, that is to say, to choose those
parameters so as to guarantee that the qualitative and
quantitative predictions of the ensuing rod theory allow for a
quick, but technically sufficient,   evaluation of the behavior of
the corresponding three-dimensional body.

Customarily, the three-dimensional theory one picks is the
treatment of the cases of flexure and torsion of the
\emph{Saint--Venant Problem} in classic linearly isotropic
elasticity. With this choice, one is led to set:
\begin{equation}\label{factors}
\mathfrak{s}_s=\frac{GA}{\chi_s}\,,\quad\mathfrak{s}_t=\frac{GJ_o}{\chi_t}\,,
\end{equation}
where $G$ is the shear modulus of the material Saint--Venant's
prismatic cylinder is comprised of, $A$ and $J_o$ are the area and
the polar moment of inertia of the cylinder's cross section, and
the dimensionless coefficients $\chi_s$ and $\chi_t$ are the
\emph{shear} and \emph{torsion factors}. Generally, these
identifications are justified by inspection of the Saint--Venant
solutions for flexure and torsion in terms of displacement
components. Each of these solutions is constructed by means of
Saint--Venant's \emph{semi-inverse method}, starting from an
educated guess about a representation for the displacement field
depending on an as-short-as-possible list of parameter functions.
Since the analytic forms of these functions depend solely on the
shape of the cross section, the same is automatically true for the
shear and torsion factors.

In textbooks of structural mechanics,  the shear factor is
interpreted as a measure of the non-uniformity in the distribution
over the cross section of Saint--Venant's cylinder of the
tangential stresses induced by the shear force; it is stated, but
as a rule not proved, that
\begin{equation}\label{chis}
\chi_s >1.
\end{equation}
As to the torsion factor, it is generally stated, and proved, that
\begin{equation}\label{chit}
\chi_t\geq 1,
\end{equation}
with equality holding if and only if the cross section is a circle
or a circular annulus. While this facts suggests that $(\chi_t
-1)$ measures the deviation from central symmetry of the cross
section, a precise interpretation is generally non attempted; at
best,  Saint-Venant's conjecture that a circular section minimizes
the twist angle among all simply-connected sections with the same
area is mentioned.

In this note we approach the issue of parameter identification as
is done in \cite{PPG}: we require that a linearly elastic rod
stores the same energy per unit length as a Saint-Venant cylinder
subject to end loads having the same force and moment resultant.
The expression we use for the rod energy is in terms of either the
shear resultant force  or the twisting moment, that for
Saint-Venant's cylinder is in terms of stress components. Both to
prove \eqref{chis} and to characterize the torsion optimality of
center-symmetric
 cross sections, we make no use of Saint-Venant's solutions in terms of displacement; the only assumptions we borrow from Saint-Venant's
 problem are those restricting the admissible external forces and the class of stress fields.

%
%
%{\color{blue}The reckoning of the shear and torsional stiffnesses
%as from Saint--Venant problem is commonly made using the
%Clapeyron's theorem on deformation working, requiring that the
%three-dimensional form be corresponding to the one-dimensional
%expression assumed by the working. Such a procedure is customarily
%carried out starting from the Saint--Venant's solution expressed
%in terms of the displacement field, providing a definition of the
%shear and torsional factors $\chi_s$ and $\chi_t$ which involves
%those kinematical parameters characterizing the solution; perhaps
%it is worth remarking that such a definition does not lead to a
%tangible feeling of the meaning of these quantities.
%
%In this work we will proceed in an other way to solve the issue.
%We first provide a new definition of the shear and torsional
%stiffnesses and factors which involves the stress components only,
%thereby making evident the physical interpretation of these
%quantities, as made in \cite{PPG}; starting from our definitions,
%we are able to prove in a simple way theirs lower bounds.
%
%In view of this purpose, we will pick from the Saint--Venant
%problem only the strictly necessary: the geometry and the
%constitutive class of the body object of our debate, the kind of
%loads which we think the body is subjected to (in particular the
%boundary conditions) and the \emph{a priori} restriction on the
%class of solutions he was looking for, commonly known as
%\emph{Saint--Venant's hypothesis}. The solution of the problem is
%not necessary for our developments, and it will not be taken into
%account.
%}

\section{The shear and torsion factors}
Given a cartesian reference frame $\{O\,; x_1, x_2, x_3\}$ and the
associated orthonormal basis $\{\eb_1, \eb_2, \eb_3\}$, we
consider a Saint--Venant's cylinder whose axis is parallel to
$\eb_3$ and whose cross section has its barycenter $O$ on the
$x_3$-axis. The vector $\xb=x_1\eb_1+x_2\eb_2$ yields the position
of a point of a typical cross section $\A$ with respect to $O$. On
denoting by $S_{ij}$ ($i,j=1,2,3$) the cartesian components of the
stress tensor,  with Saint--Venant we restrict attention to cases
when both external distance forces and external contact forces
over the cylinder's mantle are null and, moreover,
\begin{equation}\label{sv}
S_{\alpha\beta}\equiv 0\quad(\alpha,\beta=1,2).
\end{equation}
 Consequently, in order to satisfy the equilibrium equations,
not only the fields $S_{3\alpha}$ ($\alpha=1,2$) must be
independent of $x_3$ but also
\begin{equation}\label{bc}
\sb\cdot\nb=S_{3\alpha}n_\alpha=0\,,\quad \xb\in\partial\!\A\,,
\end{equation}
where $\sb=S_{3\alpha}\eb_\alpha$ is the cross-sectional traction,
 $\nb=n_\alpha\eb_\alpha$ is the normal to the mantle, and $\partial\!\A$ is the boundary curve of $\A$.

When expressed in terms of stress components, the stored-energy
density for unit volume of an isotropic material  is:
\begin{equation}\label{energy}
w(\Sb)=
%\frac{1}{2}(\Co^{-1}\Sb\cdot\Sb)=
\frac{1}{4\mu}\Big(|\Sb|^2-\frac{
\lambda}{3\lambda+2\mu}(\tr\Sb)^2\Big),
\end{equation}
where
%$\Co=\lambda\Ib\otimes\Ib+2\mu\Ib$ is the isotropic elasticity tensor with
$\lambda$ and $\mu\equiv G$ are the Lam\'e coefficients,
$|\Sb|=(S_{ij}S_{ij})^{1/2}$, and $\tr\Sb=S_{11}+S_{22}+S_{33}$;
%($\mu$), $\Ib$ the identity tensor.
under Saint-Venant's assumption \eqref{sv},  \eqref{energy}
becomes:
$$
w_{SV}=\frac{1}{2G}(S_{31}^2+S_{32}^2)+\frac{1}{2E}S_{33}^2,\quad
E:=\frac{\mu(3\lambda+2\mu)}{\lambda+\mu},
$$
where $E$ is the Young modulus. Thus, in Saint-Venant's cylinder,
the stored-energy density splits additively into two terms: the
first one is non-null if and only if the applied loads induce
cross-sectional stress, while the second one is non-null if and
only if the applied loads induce axial stress. On the other hand,
the stored-energy density per unit length of a linearly elastic
rod is:
\begin{equation}\label{stenrod}
w_s=\frac{1}{2}\frac{T^2}{\mathfrak{s}_s}
\end{equation}
when the rod is subjected to a shear force of magnitude $T$, and
is:
\begin{equation}\label{stenrodd}
w_t=\frac{1}{2}\frac{M_t^2}{\mathfrak{s}_t}
\end{equation}
when the rod is subjected to a torsion moment of magnitude $M_t$.

Following \cite{PPG}, we identify the shear-stiffness parameter by
imposing that
%
%the elastic energy per unit length for the one--dimensional model
%is by definition
%Thus we require that $w_b=\int_{\A} w_{SV}\dr A$,
%where $\dr A$ is the element of area of $\A$,
$$
\displaystyle
\frac{1}{2}\frac{T^2}{\mathfrak{s}_s}=\left(w_s=\int_{\A}
w_{SV}\dr A\right)=\frac{1}{2G}\int_{\A} (S_{31}^2+S_{32}^2)\dr
A\,,
$$
with the integrand on the right side proportional to $T^2$; hence,
\begin{equation}\label{prefacts}
\boxed{\mathfrak{s}_s:=G\frac{T^2}{\int_{\A}(S_{31}^2+S_{32}^2)\dr
A}}\,.
\end{equation}
Next, we compare this relation with the first of \eqref{factors}:
since the factor multiplying the shear modulus $G$ has the
dimension of an area, we find it natural to give $\mathfrak{s}_s$
the form \eqref{chis} by defining
\begin{equation}\label{chi_s}
\chi_s:=\frac{A\int_{\A} (S_{31}^2+S_{32}^2)\dr A}{T^2}\,,\quad
A=\int_{\A} \dr A\,.
\end{equation}

Quite similarly,  we identify the torsion-stiffness parameter by
imposing that
$$
\frac{1}{2}\frac{M_t^2}{\mathfrak{s}_t}=\frac{1}{2G}
\int_{\A}(S_{31}^2+S_{32}^2)\,\dr A\,,
$$
with the integrand on the right side proportional to $M_t^2$;
hence,
\begin{equation}\label{prefactt}
\boxed{\mathfrak{s}_t:=G\frac{M_t^2}{\int_{\A}(S_{31}^2+S_{32}^2)\,\dr
A}}\,.
\end{equation}
This time, the factor multiplying $G$ has the dimension of a
moment of inertia; to recover the second of \eqref{factors}, we
set:
\begin{equation}\label{chi_t}
\chi_t:=\frac{J_o\,\int_{\A}(S_{31}^2+S_{32}^2)\,\dr
A}{M_t^2}\,,\quad J_o=\int_{\A} \|\xb\|^2\dr A\,.\end{equation}
%
%Note that, with definitions \eqref{chi_s} and \eqref{chi_t}, both the shear factor and the torsion factor depend only on the shape of the cross section.

\section{Lower bounds}
{\color{black}In this section we prove the lower bounds
\eqref{chis} and \eqref{chit} for, respectively, $\chi_s$ and
$\chi_t$.
\subsection{$\chi_s>1.$}
Consider the following chain of inequalities:
\begin{equation}\label{ineq}
A\int_{\A} (S_{31}^2+S_{32}^2)\dr A\geq A\int_{\A} S_{32}^2\dr
A\geq \Big(\int_{\A} S_{32} \dr A \Big)^2\,.
\end{equation}
Note that the first inequality in \eqref{ineq} holds true with the
equality sign if and only if $S_{31}$ is identically null on $\A$;
and that the second, which is established by making use of
Jensen's inequality for convex functions, reduces to an equality
if and only if $S_{32}$ is identically constant on
$\A$.\footnote{A use of this second inequality in similar
circumstances is found on p. 475 of \cite{Ro}. } Next, without any
loss of generality, take the shear force parallel to $\eb_2$, so
that
\[
T=\int_{\A} S_{32} \dr A,
\]
\eqref{ineq} yields:
\[
A\int_{\A} (S_{31}^2+S_{32}^2)\dr A\geq T^2,
\]
and \eqref{chi_s} reduces to
\[
\chi_s\geq 1,
\]
with
\begin{equation}\label{cond}
\chi_s=1 \;\; \Leftrightarrow\;\; S_{31}\equiv 0\;\; \rm{and}
\;\;S_{32}\equiv{\rm const}\,.
\end{equation}
This} set of conditions on the stress field is incompatible with
the boundary condition \eqref{bc} no matter the shape of the cross
section,
 because the stress field must be continuous up to the boundary of the cross
 section itself. We conclude that the strict inequality \eqref{chis} must hold.

\subsection{$\chi_t\geq 1.$}
%
%Using the definition \eqref{chi_t}, we are going to show that
%$\chi_t\geq 1$, with the equality taking place if and only if the
%cross section is a circle.

By definition,
$$
M_t=\int_{\A} \big(x_1 S_{32}-x_2S_{31}\big)\dr A.
$$
With an application of Fubini's theorem, the denominator in
\eqref{chi_t} can be written as
\[
M_t^2  = \int_{{\A}\times{\A}} \Big( x_1
S_{32}(\xb)-x_2S_{31}(\xb)\Big) \Big( {y}_1 S_{32}({\yb})-{y}_2
S_{31}({\yb})\Big)\dr A \dr {A},
\]
or rather, equivalently, given that
\[
\int_{{\A}\times{\A}}x_1 y_2\,S_{32}(\xb)S_{31}(\yb)\dr A\dr {A}=
\int_{{\A}\times{\A}}{y}_1x_2\,S_{32}(\yb)S_{31}(\xb)\dr A\dr
{A}\,,
\]
as
\begin{equation}\label{eq:1}
M_t^2 = \int_{{\A}\times{\A}}\Big(
x_1y_1\,S_{32}(\xb)S_{32}({\yb})+x_2y_2\,S_{31}(\xb)S_{31}(\yb)
-2x_1y_2\,S_{32}(\xb)S_{31}(\yb) \Big)\dr A\dr{A}\,.
\end{equation}
An iterated use of the inequality:
\begin{equation}\label{ineq1}
\pm 2ab\leq a^2+b^2
\end{equation}
yields:
\[
\left\{\begin{array}{c}\begin{aligned}
&\int_{{\A}\times{\A}}x_1y_1\,S_{32}(\xb)S_{32}({\yb})\dr A\dr
{A}\,\leq\,
\frac{1}{2}\int_{{\A}\times{\A}}\Big(x_1^2S_{32}^2(\yb)+{y}_1^2S_{32}^2(\xb)
\Big)\dr A\dr{A}=\\
&\hspace{7cm}=\int_{{\A}\times{\A}}\Big(x_1^2S_{32}^2({\yb})
\Big)\dr A\dr {A}\,,
\end{aligned} \\\begin{aligned}
&\int_{{\A}\times{\A}}x_2y_2\,S_{31}(\xb)S_{31}(\yb)\dr A \dr
{A}\,\leq\,
\frac{1}{2}\int_{{\A}\times{\A}}\Big(x_2^2S_{31}^2(\yb)+{y}_2^2S_{31}^2(\xb)
\Big)\dr A\dr {A}=\\
&\hspace{7.cm}=\int_{{\A}\times{\A}}\Big(x_2^2S_{31}^2({\yb})
\Big)\dr A\dr {A}\,,
\end{aligned} \\\begin{aligned}
&\int_{{\A}\times{\A}}-2x_1S_{32}(\xb)\,{y}_2S_{31}(\yb)\dr A\dr
{A}\,\leq\,
\int_{{\A}\times{\A}}\Big(x_1^2S_{31}^2(\yb)+{y}_2^2S_{31}^2(\xb)\Big)\dr
A \dr {A}=\\
&\hspace{5.5cm}=\int_{{\A}\times{\A}}\Big(x_1^2S_{31}^2(\yb)+{x}_2^2S_{31}^2(\yb)\Big)\dr
A \dr {A}\,.
\end{aligned}
\end{array}\right.
\]
With this and another use of Fubini's Theorem, it follows from
\eqref{eq:1} that
$$
\begin{aligned}
M_t^2 &\leq \int_{{\A}\times{\A}}
\Big(x_1^2S_{32}^2(\yb)+x_2^2S_{31}^2(\yb)+x_1^2S_{31}^2(\yb)+x_2^2S_{32}^2(\yb)
\Big)\dr A\dr {A}=\\
&=\int_{{\A}\times{\A}}\|\xb\|^2\big(S_{31}^2(\yb)+S_{32}^2(\yb)\big)\dr
A \dr {A}= \Big(\int_{\A}\|\xb\|^2\dr
A\Big)\,\int_{\A}\big(S_{31}^2(\xb)+S_{32}^2(\xb)\big)\dr A\,=\\
&=J_0\int_{\A}(S_{31}^2+S_{32}^2)\dr A\,,
\end{aligned}
$$
which is tantamount to having from \eqref{chi_t} that $\chi_t\geq
1$ whatever the shape of the cross section.\footnote{A reviewer of
our paper remarked that the inequality $M_t^2 \leq
J_0\int_{\A}(S_{31}^2+S_{32}^2)\dr A$ can be proved by a
straightforward application of the Cauchy-Bunyakovsky-Schwarz
inequality.}

Now, given that equality holds in \eqref{ineq1} if and only if
$a=\mp b$, $\chi_t= 1$ if and only if all of the following
conditions are identically satisfied in $\A\times\A$:
\begin{equation}
x_1S_{32}(\yb)={y}_1S_{32}(\xb),\quad
x_2S_{31}(\yb)={y}_2S_{31}(\xb),\quad
x_1S_{31}(\yb)=-{y}_2S_{32}(\xb);
\end{equation}
the first two are implied by the last, which can be written as:
$$\frac{S_{32}(\xb)}{x_1}=-\frac{S_{31}(\yb)}{{y}_2}=c\,,
$$
or rather, equivalently,
\begin{equation}\label{form}
\sb=c\,(-x_2\eb_1+x_1\eb_2)=c\,\eb_3\times\xb,
\end{equation}
with $c$ a constant. Hence, the cross-sectional traction $\sb$
must be orthogonal to the position vector $\xb$ all over $\A$ up
to the boundary, where on the other hand it has to be orthogonal
to the normal $\nb$ to satisfy \eqref{bc}. Because of the assumed
continuity of the stress field up to the boundary, $\xb$ must then
be parallel to $\nb$ all over $\partial\!\A$, which is possible if
and only if ${\A}$ is a circle or a circular annulus. Thus, as
anticipated in the introduction, $(\chi_t-1)$ quantifies the
reduction in torsional stiffness due to the defect in polar
symmetry of the cross section and the accompanying deviation from
the form \eqref{form} of the cross-sectional traction.

\section{Final remarks}
\vspace{0.3cm} \noindent\emph{Remark 1} One can define the
\emph{extension} and \emph{bending stiffnesses} $\mathfrak{s}_e$
and $\mathfrak{s}_b$ and the relative factors $\chi_e$ and
$\chi_b$ of a rod by posing, respectively,
\begin{equation}\label{prefacte}
\boxed{\frac{1}{2}\frac{N^2}{\mathfrak{s}_e}:=\frac{1}{2E}
\int_{\A}S_{33}^2\,\dr A}\,,\quad \chi_e:=\frac{A\int_{\A}
S_{33}^2\dr A}{N^2}
\end{equation}
for
\[
N=\int_{\A} S_{33}\dr A,
\]
 and
\begin{equation}\label{prefactb}
\boxed{\frac{1}{2}\frac{M^2}{\mathfrak{s}_b}:=\frac{1}{2E}
\int_{\A}S_{33}^2\,\dr A}\,,\quad \chi_b:=\frac{J\int_{\A}
S_{33}^2\dr A}{M^2}
\end{equation}
for
\[
M=\int_{\A} x_{2}S_{33}\dr A, \quad J=\int_{\A} x_{2}^2\dr A\,.
\]
To prove that
\begin{equation}\label{chieb}
\chi_e\geq1 \quad\textrm{and}\quad \chi_b\geq 1
\end{equation}
is left as an exercise for the interested reader. We note that
equality is realized in relations \eqref{chieb} whatever the shape
of the cross section $\A$ if and only if, respectively, the field
$S_{33}$ is constant-valued or linear in $x_2$ over $\A$, just as
it happens to be in Saint-Venant's cases of normal force and pure
bending.

\vspace{0.3cm} \noindent\emph{Remark 2} The recipe for parameter
identification embodied in our definitions \eqref{prefacts},
\eqref{prefactt}, $\eqref{prefacte}_1$, and $\eqref{prefactb}_1$
for rod stiffnesses is applicable as such to any one-dimensional
counterpart of a three-dimensional rod-like material body, no
matter its axis were straight, its cross section constant or its
mechanical response spatially uniform, as is the case for
Saint-Venant's prismatic cylinder; while linearity in the elastic
response is crucial, isotropy is not: all those definitions make
sense also for \emph{transverse isotropy} with respect to the
axial direction.

\end{document}